\documentclass[prl,aps,twocolumn,showpacs]{revtex4}

\usepackage{epsfig}
\usepackage{amssymb}
\usepackage{graphicx}

\begin{document}

\sloppy

\title{Generation and purification of maximally-entangled atomic states in optical
cavities}

\author{P. Lougovski$^{1}$, E. Solano$^{1,2}$, and H. Walther$^{1}$}

\affiliation{$^1$Max-Planck-Institut f{\"u}r Quantenoptik, Hans-Kopfermann-Strasse 1,
D-85748 Garching, Germany\\
$^2$Secci\'{o}n F\'{\i}sica, Departamento de Ciencias, Pontificia
Universidad Cat\'{o}lica del Per\'{u}, Apartado 1761, Lima, Peru }

\begin{abstract}
We present a probabilistic scheme for generating and purifying
maximally-entangled states of two atoms inside an optical cavity
via no-photon detection at the cavity output, where ideal
detectors are not required. The intermediate mixed states can be
continuously purified so as to violate Bell inequalities in a
parametrized manner. The scheme relies on an additional
strong-driving field that realizes, atypically, simultaneous
Jaynes-Cummings and anti-Jaynes-Cummings interactions.
\end{abstract}

\vspace*{-0.5cm}

\maketitle

Entanglement, first considered by
Schr{\"o}dinger~\cite{Schrodinger}, is recognized nowadays as a
cornerstone in the fundamentals of quantum physics and as a source
of diverse applications in quantum information and
computation~\cite{Nielsen}. In particular, entangled states of
discrete systems, such as two or more qubits, play an important
role in testing fundamental properties of quantum theory. They
allow one, for instance, to prove the nonlocal character of
quantum mechanics versus local hidden-variable theories.
Maximally-entangled states of two-qubit systems have already been
produced experimentally in photonic systems~\cite{Kwiat} and in
the internal degrees of freedom of atoms interacting with a
microwave cavity~\cite{Walther,Haroche}. In the case of trapped
ions~\cite{ionsNist,ionsInnsbruck}, maximally entangled states
have been created through the manipulation of their collective
motion, but cavity QED devices are needed for transferring the
stored information. Despite the diverse and recent theoretical
proposals, see Refs.~\cite{Beige,Molmer,Marr} and references
therein, generation of maximally-entangled states of two atoms
inside an optical cavity has not yet been accomplished in the lab.
The relevance of this achievement strongly relies on the
possibility of using atoms in optical cavities as quantum
networks~\cite{CiracZollerNetworks}, where quantum processing
could take place among the entangled atoms and quantum information
could be distributed among distant
cavities~\cite{Feng,Molmer1,Kimble,Plenio,Kraus}. Here, we propose
a scheme that addresses most of the problems of a realistic model
for entangling two atoms inside an optical cavity: dissipative
processes, atomic localization, detection efficiency and purity of
the generated entangled state.

We consider two identical three-level atoms in
$\Lambda$-configuration placed inside an optical cavity, see Fig.
1(a), where the allowed transitions
$|c\rangle\leftrightarrow|g\rangle$ and
$|e\rangle\leftrightarrow|g\rangle$ are excited off-resonantly by
laser fields and a cavity mode, see Fig. 1(b). The metastable
states $|g\rangle$ and $|e\rangle$ are resonantly coupled through
level $| c \rangle$ by two effective interactions, one stemming
from a laser field and the cavity mode and the other from two
additional laser fields. The different frequency detunings,
$\Delta$ and $\Delta'$, of these two $\Lambda$-processes prevent
the system from undesired transitions. We assume that both atoms
couple to the cavity mode with similar strength $g$, taken as real
as all other coupling strengths $\{\Omega, \Omega'_1, \Omega'_2
\}$ for the sake of simplicity. Then, the Hamiltonian for the
system can be written as
\begin{eqnarray}
\label{hamiltonian}H & = & \hbar\omega_{e}\sum^2_{j=1}|e_{j}\rangle
\langle e_{j}| + \hbar\omega_{c}\sum^2_{j=1}|c_{j}\rangle\langle c_{j}|
+ \hbar\omega_{f}a^{\dagger}a   \nonumber \\
& & + \hbar g (a^{\dagger}\sum^2_{j=1}|e_{j}\rangle\langle c_{j}|
+ a\sum^2_{j=1}|c_{j}\rangle\langle e_{j}|)   \nonumber\\
& & + \hbar\Omega (e^{-i(\omega_{c}-\Delta)t}
\sum^2_{j=1}|c_{j}\rangle\langle g_{j}| + h.c.)  \\
& & + \hbar\Omega'_{2}(e^{-i(\omega_{c} - \omega_{e}
-\Delta')t}\sum^2_{j=1}|c_{j}\rangle\langle e_{j}| + h.c.) \nonumber\\
 & & + \hbar\Omega'_1 (e^{-i(\omega_{c}-\Delta')t}
 \sum^2_{j=1}|c_{j}\rangle\langle g_{j}| + h.c.) . \nonumber
\end{eqnarray}
Here, $\omega_{c}$ and $\omega_{e}$ are the Bohr frequencies
associated with the transitions
$|c\rangle\leftrightarrow|g\rangle$ and
$|e\rangle\leftrightarrow|g\rangle$, respectively, while
$\omega_f$ is the frequency of the cavity mode and $a$
($a^{\dagger}$) the associated annihilation (creation) operator.
To eliminate level $| c \rangle$ adiabatically, so as to discard
spontaneous emission from our model, we require
\begin{equation}
\label{adiabatic}\{ \frac{\Omega}{\Delta}, \frac{g}{\Delta},
\frac{\Omega'_{1}} {\Delta'}, \frac{\Omega'_{2}}{\Delta'} \} \ll 1
.
\end{equation}
In addition, and in order to avoid undesired atomic transitions,
we need the following rotating-wave-approximation (RWA)
inequalities
\begin{equation}
\label{RWA}\Delta-\Delta'\gg \{ \frac{\Omega
\Omega'_{2}}{\Delta'}, \frac{\Omega \Omega^{\prime}_{1}}{\Delta'},
\frac{\Omega'_{2}g}{\Delta'}, \frac{\Omega^{\prime}_{1}g}{\Delta'}
\} ,
\end{equation}
turning Eq.~(\ref{hamiltonian}) into the effective Hamiltonian
\begin{eqnarray}
\label{effect}H_{\rm eff} = - \hbar g_{\rm eff}
\sum^2_{j=1}(a^{\dagger}\sigma^{\dagger}_{j} + a\sigma_{j}) -
\hbar \Omega'_{\rm eff}\sum^2_{j=1} (\sigma^{\dagger}_{j} +
\sigma_{j}),
\end{eqnarray}
where $g_{\rm eff} \equiv \Omega g / \Delta$, $\Omega'_{\rm eff}
\equiv \Omega'_1 \Omega'_2 / \Delta'$, $\sigma^{\dagger}_{j} =
|e_{j}\rangle\langle g_{j}|$ and $\sigma_{j}=|g_{j}\rangle\langle
e_{j}|$. In Eq.~(\ref{effect}), AC Stark shifts are assumed to be
corrected by retuning the laser frequencies~\cite{BiswasAgarwal}.

\begin{figure}[!ht]
\hspace*{-7cm} \large{ a) } \vspace*{-0.6cm}
\begin{center}
\includegraphics[width=50mm]{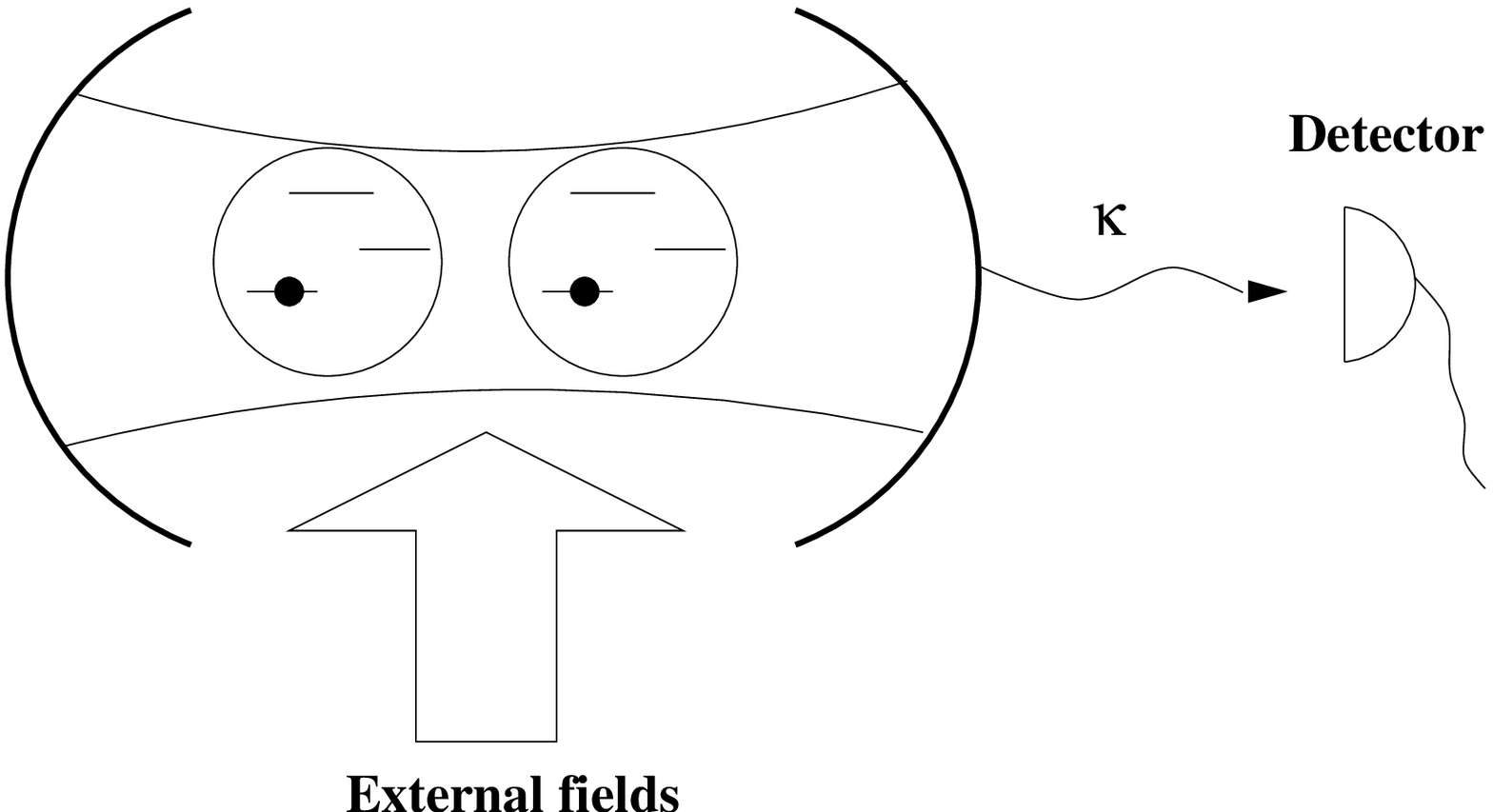}
\end{center}
\hspace*{-7cm} \large{ b) } \vspace*{-0.7cm}
\begin{center}
\includegraphics[width=40mm]{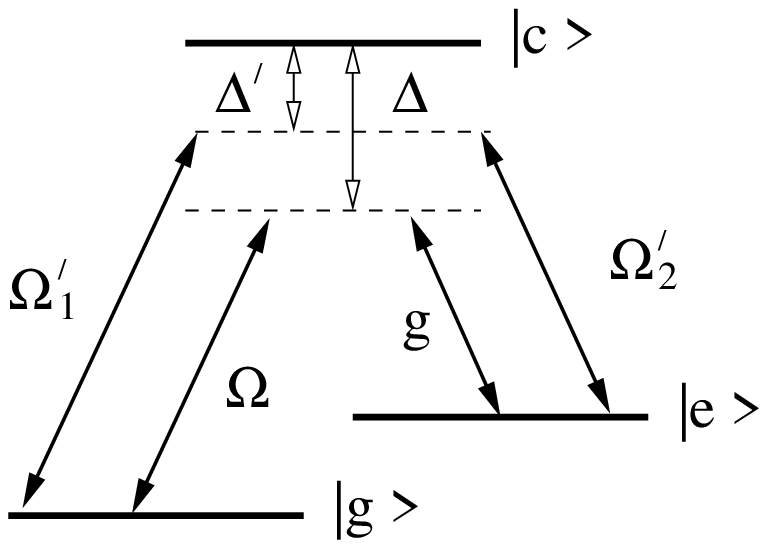}
\caption{\label{setup} (a) Two three-level atoms inside an optical
cavity, (b) scheme of the atomic energy levels and the exciting
fields for each three-level atom. $\Delta$ and $\Delta'$ are
frequency detunings and $g$, $\Omega$, $\Omega'_1$ and $\Omega'_2$
the respective coupling strengths.} \vspace*{-0.5cm}
\end{center}
\end{figure}
Eq.~(\ref{effect}) shows the effective coupling of metastable
states $| g_j \rangle$ and $| e_j \rangle$ with the cavity mode
(anti-Jaynes-Cummings) and with a classical external driving.

We now consider the external strong-driving regime, where
$\Omega'_{\rm eff} \gg g_{\rm eff}$. In this way, and going to an
interaction picture with respect to the (effective) external
driving term, the Hamiltonian of Eq.~(\ref{effect}) can be written
as
\begin{eqnarray}
\label{key}H^{\rm int}_{\rm eff} & = & - \hbar \frac{ g_{\rm eff}}
{2} (a^{\dagger} + a)\sum^2_{j=1}(\sigma^{\dagger}_{j} +
\sigma_{j}) .
\end{eqnarray}
A similar Hamiltonian was obtained in Ref.~\cite{Kike} for the
case of $N$ two-level Rydberg atoms interacting with a microwave
cavity and a strong external field, yielding a wide family of
multipartite entangled N-atom-cavity states. Here, we have shown
that a similar effective Hamiltonian can be realized in the
optical domain but, in contrast to the microwave regime, no
atom-field entanglement is expected to survive long enough for
practical purposes due to the comparatively lower achievable
ratios $g_{\rm eff}/\kappa$ ($\kappa$ being the cavity decay
rate). Nevertheless, it is possible to design strategies for using
this faster dissipation process to produce entanglement in the
atomic degrees of freedom. We will describe a scheme for the case
of two atoms whose initial state, at the time $t=0$, is the
atom-field ground state $|gg\rangle | 0 \rangle \equiv
|g_{1}\rangle\otimes|g_{2}\rangle \otimes| 0 \rangle$. The
resulting atom-field state at $t=\tau$ is then
\begin{eqnarray}
\label{state} | \Psi (\tau) \rangle & = & \frac{1}{2} | ++
\rangle| 2 \alpha ({\tau}) \rangle +
\frac{1}{2}|--\rangle|- 2 \alpha ({\tau}) \rangle \nonumber \\
& & + \frac{1}{\sqrt{2}}| \Psi^{+} \rangle|0\rangle,
\end{eqnarray}
where
\begin{eqnarray}
|++\rangle \!\! &  \equiv & \!\! |+_{1}\rangle|+_{2}\rangle =
\frac{1}{\sqrt{2}}(|g_{1}\rangle
+ |e_{1}\rangle) \times \frac{1}{\sqrt{2}} (|g_{2}\rangle + |e_{2}\rangle), \nonumber \\
|--\rangle \!\! &  \equiv & \!\! |-_{1}\rangle|-_{2}\rangle =
\frac{1}{\sqrt{2}}(|g_{1}\rangle - |e_{1}\rangle) \times
\frac{1}{\sqrt{2}} (|g_{2}\rangle - |e_{2}\rangle), \nonumber \\
\end{eqnarray}
such that the maximally-entangled state
\begin{eqnarray}
\label{eprwavefunction} | \Psi^{+} \rangle =
\frac{1}{\sqrt{2}}(|-_{1}\rangle|+_{2}\rangle +
|+_{1}\rangle|-_{2}\rangle) ,
\end{eqnarray}
and the amplitude of the coherent states generated is
\begin{eqnarray}
\label{amplitude}
\alpha (\tau) = i\hspace*{0.05cm}\frac{g_{\rm
eff}\tau}{2} .
\end{eqnarray}
We observe that the atomic states $| -_1 +_2 \rangle$ and $| +_1
-_2 \rangle$ of Eq.~(\ref{eprwavefunction}) are eigenstates of the
collective operator $\sigma_x = \sigma_{x,1} + \sigma_{x,2} =
\sum^2_{j=1}(\sigma^{\dagger}_{j} + \sigma_{j})$ with eigenvalues
equal to zero. This fact explains, following Eqs.~(\ref{key}) and
(\ref{state}), the persistent correlation of the vacuum field
state with the atomic state $| \Psi^{+} \rangle$ through the whole
unitary evolution. These states are called dark states in the
literature~\cite{Pellizzari}. Note that the size of the coherent
state, estimated by the amplitude $\alpha$ in
Eq.~(\ref{amplitude}), is proportional to the time $\tau$ of the
unitary process described by the effective Hamiltonian of
Eq.~(\ref{key}). If $| \alpha |$ is large enough ($| \alpha | \ge
2$), such that we can consider the states $| - \alpha \rangle$, $|
0 \rangle$ and $| \alpha \rangle$ as mutually orthogonal, then a
measurement of the vacuum field in the atom-field state of
Eq.~(\ref{state}) would project the atomic state onto $| \Psi^{+}
\rangle$ with probability $1/2$. In consequence, measuring a
zero-photon state leaking the cavity mode, as sketched in
Fig.~1(a), would be enough for producing an atomic $| \Psi^{+}
\rangle$ state with high fidelity. However, one would require
(unavailable) detectors with high efficiency and the (already
available) strong-coupling regime of optical cavities, see
Refs.~\cite{Rempe,Kimble1}.

It is possible to extend our method to less demanding regimes and
more realistic conditions, involving field damping, weak-coupling
regime and finite efficiency detection, without increasing the
complexity of the experimental requirements. We then write down a
master equation describing the atom-field dynamics
\begin{eqnarray}
\label{me}\dot{\rho}_{at-f} & = & - \frac{i}{\hbar}[H^{\rm
int}_{\rm eff}, \rho_{at-f}] + \mathcal{L}\rho_{at-f},
\end{eqnarray}
where the (field) dissipative term is described by
\begin{eqnarray*}
\mathcal{L}\rho_{at-f} & = &
-\frac{\kappa}{2}(a^{\dagger}a\rho_{at-f} -
2a\rho_{at-f}a^{\dagger} + \rho_{at-f}a^{\dagger}a).
\end{eqnarray*}
The master equation in Eq.~(\ref{me}) can be solved analytically
by means of phase-space techniques~\cite{solvablemasterequation},
representing an unusual case of a solvable master equation
involving coherent driving and dissipation. When the atom-field
system is prepared initially in its ground state, the steady-state
solution of the master equation reads
\begin{eqnarray}
\label{bs}\rho^{ss}_{at-f} & = & \frac{1}{4}|++\rangle\langle ++|
\otimes|2\tilde{\alpha}\rangle\langle2\tilde{\alpha}| \nonumber \\
 & + & \frac{1}{4}|--\rangle\langle --|\otimes|-2\tilde\alpha\rangle
 \langle-2\tilde{\alpha}| \\
 & + & \frac{1}{2}|\Psi^{+} \rangle\langle \Psi^{+} |\otimes|0\rangle\langle0|
 \nonumber,
\end{eqnarray}
with $\tilde{\alpha} = i\frac{g_{\rm eff}}{\kappa}$. In contrast
to the pure state of Eq.~(\ref{state}), the steady state of the
atom-cavity system is a mixed state with no quantum correlation
between the atoms and the field. The remarkable feature of the
state in Eq.~(\ref{bs}) is that the atomic state $| \Psi^{+}
\rangle$ is still correlated with the vacuum of the cavity field.
Henceforth, if one performs a first no-photon measurement of the
cavity field in the steady state, the projected (normalized)
atomic density operator is
\begin{eqnarray}
\label{atom}\rho^{ss}_{at} & = & \frac{e^{-|2\tilde{\alpha}|^{2}}}{1 +
e^{-|2\tilde{\alpha}|^{2}}}(|++\rangle\langle ++| +
|--\rangle\langle --|) \nonumber \\
 & + & \frac{1}{1 + e^{-|2\tilde{\alpha}|^{2}}}| \Psi^{+} \rangle\langle \Psi^{+} |.
\end{eqnarray}
When $| \tilde{\alpha} | > 1$, in the strong-coupling
regime~\cite{Rempe, Kimble1}, the condition
$|2\tilde{\alpha}|^{2}\gg 1$ is automatically fulfilled and
Eq.~(\ref{atom}) reduces to the maximally entangled atomic state
\begin{eqnarray}
\label{epr}\rho^{ss}_{at} & = & | \Psi^{+} \rangle\langle \Psi^{+}
|.
\end{eqnarray}
 with unity fidelity and success probability $1/2$.

If $| \tilde{\alpha} | < 1$, in the weak-coupling regime, the
desired atomic state of Eq.(\ref{epr}) will be contaminated by
other contributions as shown in Eq.~(\ref{atom}). In this case, we
are still able to develop a protocol which purifies the atomic
state $| \Psi^{+} \rangle$ via a successive application of the
same scheme. We repeat our procedure, shining a similar laser
system on our new initial atom-cavity state
\begin{equation}
\label{second}\rho_{at-f} =
\rho^{ss}_{at}\otimes|0\rangle\langle0|
\end{equation}
until it reaches a new steady state, followed by a measure of a
no-photon event with a certain finite probability. By repeating
this sequence of steps $N$ times, we arrive at the projected
atomic density operator
\begin{eqnarray}
\label{atomN}\rho^{ss}_{at}(N) & = & \frac{e^{-N|2\tilde{\alpha}|^{2}}}
{1 + 2e^{-N|2\tilde{\alpha}|^{2}}}(|++\rangle\langle ++| + |--\rangle
\langle --|) \nonumber \\
& + & \frac{1}{1 + 2e^{-N|2\tilde{\alpha}|^{2}}}| \Psi^{+}
\rangle\langle \Psi^{+} |.
\end{eqnarray}
For a given $| \tilde{\alpha} |$, even in the weak-coupling
regime, we can always choose a number of repetitions $N$ such that
$e^{-N|2\tilde{\alpha}|^{2}} \!\! = \! 0$, warranting a highly
pure atomic state $| \Psi^{+} \rangle$. In this case, the fidelity
of the state $| \Psi^{+} \rangle$ is given by
\begin{eqnarray}
\label{fidelity} \!\!\!\!\!\!\!\! F(N) & = & \langle \Psi^{+}
|\rho^{ss}_{at}(N)| \Psi^{+} \rangle = \frac{1}{1 +
2e^{-N|2\tilde{\alpha}|^{2}}},
\end{eqnarray}
with success probability
\begin{eqnarray}
\label{sp}P_{suc} & = & \frac{1}{2}\frac{1}{1 +
e^{-|2\tilde{\alpha} |^{2}}}\prod^{N}_{m=2}\frac{1}{1 +
2e^{-m|2\tilde{\alpha}|^{2}}}.
\end{eqnarray}
Note that even for $| \tilde{\alpha} | \sim 1$, $N = 1,2$,
fidelity $F(N) \sim 1$.

Even in the weak-coupling regime, $| \tilde{\alpha} | < 1$, the
vacuum state $| 0 \rangle$ can be neatly distinguished from
coherent states $| \pm \tilde{\alpha} \rangle$. This is a natural
consequence of the fact that if one photon is emitted by measuring
$| \pm \tilde{\alpha} \rangle$, but possibly not detected due to
dark counts or any other noisy effect, there will be further
emission of photons due to the continuous laser pumping
mechanism~\cite{Plenio2}. In other words, our measurement is split
in two distinct outcomes, or we have repeated opportunities of
detecting no photon, and the protocol continues, or we have
repeated opportunities of measuring single photons, which means
that we should start again. In this way, we are able to overcome
the detection efficiency problem.

It is not necessary to wait until the steady state of
Eq.~(\ref{bs}) is reached to realize this protocol. For example,
one could implement the following variation: i) production of
state in Eq.~(\ref{state}) followed by disconnection of lasers,
ii) photo-detection (unaffected by decay process), iii) if
no-photon was detected then back to (i) until desired fidelity is
reached, iv) if a photon is detected then restart the protocol.
This variation minimizes the time in which the lasers are on and
assures no influence of spontaneous-emission effects at all.

\begin{figure}[!t]
\begin{center}
\includegraphics[scale=0.55]{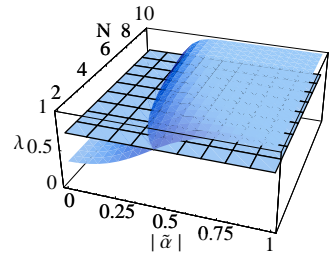}
\vspace*{-0.5cm} \caption{\label{bell} Violation of the Bell
inequality.}\vspace*{-0.5cm}
\end{center}
\end{figure}

It is important to estimate the influence of atomic localization
on the generation of entanglement, as long as our method relies
strongly on the production of dark states~\cite{Pellizzari}. The
fidelity of the purified dark state in Eq.~(\ref{epr}) follows $F
= 1 / 1 + \epsilon^2$. Here, $\epsilon \equiv \delta g / \kappa$,
where $\delta g$ is the differential variation of the atom-field
coupling due to the differential variation in the localization of
the two atoms. The parameter $\epsilon$ changes very slowly with
possible errors in the atomic locations, due to their intrinsic
cosine dependence. For realistic parameters,
see~\cite{CQEDionsMPQ}, a maximal localization error of $10 \%$ of
an optical wavelenght yields $\epsilon \approx 0.1$. This implies
a fidelity $F > 0.99$, showing the robustness of our proposal.

It was proved~\cite{Gisintheory,Gisinexperiment} that a family of
mixed states, similar to that of Eq.~(\ref{atomN}),
\begin{eqnarray}
\label{GisinState}\rho & = & \frac{1-\lambda}{2}(|++\rangle\langle ++|
+ |--\rangle\langle --|) \nonumber \\
 & + & \lambda| \Psi^{+} \rangle\langle \Psi^{+} | ,
\end{eqnarray}
violates the Bell inequality if and only if
$\lambda>\frac{1}{\sqrt{2}}$. In our case $\lambda$ is a function
of two parameters, $|\tilde{\alpha}|$ and $N$. In Fig. 2, we plot
the surface $\lambda(|\tilde{\alpha}|,N)$ cut by a plane
corresponding to the boundary value $\frac{1}{\sqrt{2}}$. There,
we can clearly see that it is possible to cross the threshold
parameter $\lambda = 1 / \sqrt{2}$ by increasing the number $N$ of
repetitions and the amplitude $\tilde{\alpha}$ in the proposed
scheme. This transition shows a local parametrized evolution from
a mixture (with classical correlations) to a maximally-entangled
atomic state (with quantum correlations), as was discussed in
Refs.~\cite{Gisintheory,Gisinexperiment}.

Finally, we illustrate our protocol with a variant for direct
generation of another maximally-entangled state. We assume that in
the previous scheme, see Fig.~1, the coupling strength is $\Omega
= 0$ and the detuning frequencies are
$\Delta_{1}=\Delta_{2}=\Delta$. Suppose now that the atoms couple
differently to the cavity mode in such a way that the coupling
strengths have the same absolute values but the opposite phases
($g_{1}=|g|, g_{2}=-|g|$). In this case, the adiabatic elimination
conditions of Eq.~(\ref{adiabatic}) reduce to
\begin{equation}
\label{newad} \{ \frac{\Omega'_{1}}{\Delta},
\frac{\Omega'_{2}}{\Delta},
 \frac{|g|}{\Delta} \} \ll 1 .
\end{equation}
Therefore, the effective Hamiltonian in the interaction picture,
after imposing the strong-(external)driving regime $ \Omega'_{\rm
eff} \ll |g'_{\rm eff}|$, with $g'_{\rm eff} \equiv \Omega'_1 g /
\Delta$, reads
\begin{eqnarray}
\label{nkey}\tilde{H}^{\rm int}_{\rm eff} & = & \hbar
\frac{g'_{\rm eff}}{2} (a^{\dagger} +
a)\sum^2_{j=1}(-1)^{j}(\sigma^{\dagger}_{j} + \sigma_{j}).
\end{eqnarray}
The Hamiltonian of Eq.~(\ref{nkey}) is slightly different from the
Hamiltonian of Eq.~(\ref{key}) and, when substituted in
Eq.~(\ref{me}), yields the atom-field steady state
\begin{eqnarray}
\label{ns}\rho^{ss}_{at-f} & = & \frac{1}{4}|-+\rangle\langle +-|
\otimes|2\beta\rangle\langle2\beta| \nonumber \\
 & + & \frac{1}{4}|+-\rangle\langle -+|\otimes|
 -2\beta\rangle\langle-2\beta| \\
 & + & \frac{1}{2}|\Phi^{+}\rangle\langle \Phi^{+}|
 \otimes|0\rangle\langle0| \nonumber,
\end{eqnarray}
where $|\beta\rangle$, $|-\beta\rangle$ are coherent states with
$|\beta| = \frac{g'_{\rm eff}}{\kappa}$ and
\begin{eqnarray}
|\Phi^{+}\rangle & = &
\frac{1}{\sqrt{2}}(|+_{1}\rangle|+_{2}\rangle +
|-_{1}\rangle|-_{2}\rangle) \nonumber
\end{eqnarray}
is another maximally-entangled Bell state. Purification of the
state $|\Phi^{+}\rangle$, out of the steady state in
Eq.~(\ref{ns}), can be done by following steps similar to the ones
before.

The coupling of atoms to the cavity mode, with similar or opposite
phase, can be achieved with ion traps~\cite{CQEDionsMPQ}.

In conclusion, we have proposed a protocol for generating and
purifying maximally-entangled states in the internal degrees of
freedom of two atoms inside an optical cavity. Our protocol is of
a quantum-non-demolition kind, where successive no-photon
detection of the cavity field projects and purifies sequentially
the desired atomic state. In contrast to recent proposals, our
scheme produces atomic Bell states as a steady state of the
two-atom-field interaction in correlation with the vacuum field
state, which makes our method robust to decoherence. It combines a
reasonably high success probability ($\sim 1/2$) with very high
fidelity ($\sim 1$) for a wide range of parameters. The proposed
scheme does not rely on a high detection efficiency as long as it
is based on discrimination, through projection, between a vacuum
field state (zero detector clicks) and orthogonal coherent states
(necessarily more than one detector click). Furthermore, the
procedure described above not only allows efficient creation of
maximally-entangled states in two-atom states inside optical
cavities under realistic conditions, but also suggests strategies
for purifying chosen entangled states with known contamination
states.

P. L. acknowledges financial support from the Bayerisches
Staatsministerium f\"ur Wissenschaft, Forschung und Kunst in the
frame of the Information Highway Project and E. S. from the EU
through the RESQ (Resources for Quantum Information) project.


\begin{thebibliography}{99}
\bibitem{solvablemasterequation} The analytical solution will be presented elsewhere.
\end{thebibliography}
\end{document}